 \documentclass[12pt,preprint]{aastex}

\newcommand{\fex}{[Fe~X]$\lambda$6374}
\newcommand{\fexiv}{[Fe~XIV]$\lambda$5303}
\newcommand{\bd}{BD+30$^\circ$3639}

\slugcomment{}

\shorttitle{\fexiv\ in PNe}

\shortauthors{Georgiev et al.}

\begin{document}


\title{Iron depletion in the hot bubbles in planetary nebulae \footnote{Based upon observations obtained at the Observatorio Astronmico Nacional in San Pedro M\'artir, Baja California, Mexico}\ $^\mathrm ,$\footnote{Based on observations obtained with XMM-Newton, an ESA science mission with instruments and contributions directly funded by ESA Member States and NASA}}


\author{Leonid N. Georgiev}
\affil{Instituto de Astronom\'\i a, UNAM, Apartado Postal 70-254,
CD Universitaria, CP 04510, M\'exico, D.F., M\'exico}
\email{georgiev@astroscu.unam.mx}

\author{Michael G. Richer}
\affil{Observatorio Astron\'omico Nacional, Instituto de Astronom\'\i a,
UNAM, P.O. Box 439027, San Diego, CA 92143, USA}
\email{richer@astrosen.unam.mx}

\author{Anabel Arrieta}
\affil{Universidad Iberoamericana, Departamento de F\'\i sica y
Matem\'aticas, Av. Prolongaci\'on Paseo de la Reforma 880, Lomas de
Santa Fe, CP 01210, M\'exico, D.F., M\'exico}
\email{anabel.arrieta@uia.mx}

\and

\author{Svetozar A. Zhekov\altaffilmark{3}}
\affil{JILA, University of Colorado, Boulder, CO 80309-0440, USA}
\email{zhekovs@colorado.edu}

\altaffiltext{3}{On leave from Space Research Institute, Sofia, Bulgaria}



\begin{abstract}

We have searched for the emission from \fex\ and \fexiv\ that is expected from the gas emitting 
in diffuse X-rays in \bd, NGC 6543, NGC 7009, and NGC 7027.  Neither line was
detected in any object.  Models that fit the X-ray spectra of these objects indicate that 
the \fex\ emission should be below our detection thresholds, but the predicted \fexiv\ 
emission exceeds our observed upper limits (one sigma) by factors of at least 3.5 to 12.  
The best explanation for the absence of \fexiv\ is that the X-ray plasma is depleted in iron.  
In principle, this result provides a clear chemical signature that may be used to determine 
the origin of the X-ray gas in either the nebular gas or the stellar wind.  At present, though 
various lines of evidence appear to favour a nebular origin, the lack of atmospheric and 
nebular iron abundances in the objects studied here precludes a definitive conclusion.  

\end{abstract}




\section{Introduction}

\begin{deluxetable}{lccrll}
\tablecolumns{6}
\tablecaption{Log of observations}
\label{obslog}
\tablehead{ \colhead{Date (UT)} & \colhead{Object} & \colhead{Slit PA$^\mathrm {a}$} 
& \colhead{Slit size} & \colhead{Slit center} & \colhead{Exposures} }
\startdata
26 May 2004 & NGC 6543 & 10$^\circ$ & $  2\arcsec \times 26\arcsec$ &  the star         & $7\times 900$s   \\
27 May 2004 & NGC 6543 & 43$^\circ$ & $  2\arcsec \times 26\arcsec$ &  the star         & $11 \times 900$s \\
30 May 2004 & \bd      & 0$^\circ$  & $3.3\arcsec \times 13\arcsec$ & 1$\arcsec$ east from the star        & $5\times 1800$s  \\
 2 Aug 2004 & NGC 7027 & 138$^\circ$& $3.3\arcsec \times 26\arcsec$ &  the bright spot  & $5\times 600$s   \\
 3 Aug 2004 & NGC 7009 & 72$^\circ$ & $  2\arcsec \times 26\arcsec$ &  the star         & $12\times 900$s  \\
 4 Aug 2004 & NGC 7009 & 162$^\circ$& $  2\arcsec \times 26\arcsec$ &  the star         & $11\times 900$s  \\
\enddata
\begin{list}{}{}
\item[$^{\mathrm a}$] The positions are shown in Fig.~\ref{all_obj}.  The position angle follows astronomical
convention, measured north through east.
\end{list}
\end{deluxetable}

Planetary nebulae, one of the final evolutionary phases for low- and intermediate-mass stars, 
are believed to be formed as a result of the interaction of the fast wind expelled by the central 
star with a slow wind expelled at the AGB phase \citep{kwoketal1978}. The strong interaction between 
these two winds produces a bubble of shock-heated gas whose temperature, according to the fast wind 
velocity, is expected to be several times $10^7$\,K. The first diffuse X-ray emission observed from 
planetary nebulae by the ROSAT satellite (ROSAT detected only {\it spectral} evidence for nebular emission
from PNe but did not spatially resolve it), the Chandra X-ray Observatory, 
and XMM-Newton was very surprising because the observed temperatures of the X-ray-emitting gas were 
$\sim 1-3\times 10^6$K, up to an order of magnitude lower than expected \citep{kreysingetal1992,
guerreroetal2000,kastneretal2000,chuetal2001,kastneretal2001,guerreroetal2002}. Since then, diffuse 
X-rays have been observed in Mz 3, Hen 3-1475, and NGC 2392 \citep[respectively]{kastneretal2003,sahaietal2003,guerreroetal2005}. 
Of these, the first two have temperatures as high or higher than previously observed while the 
X-ray emission in NGC 2392 is characterized by a temperature of $\sim 2\times 10^6$\,K, which is 
approximately the temperature expected given the current wind velocity of its central star.

The generally low temperature of the gas emitting X-rays raises
the problem of cooling this gas to the observed temperatures.
Since the hot gas is in contact with the ``cold" optical shell,
thermal conduction may play an important role and may lower the
gas temperature of the hot bubble, thereby solving the \lq\lq
softness" problem in the X-ray spectra of PNe
\citep{soker1994,zhekovperinotto1998}. \citet{sokerkastner2003}
discussed several \lq\lq cooling" mechanisms ranging from thermal
conduction and fluid mixing of X-ray-emitting and nebular material
to the effects of adiabatic expansion, winds, and jets.  Other
processes, such as the influence of magnetic fields, e.g.,
\citet{rozyczkafranco1996}, have yet to be investigated. The
cooling mechanism could be related to the gas distribution.
\citet{kastneretal2002} present high resolution images that appear
to indicate that the X-ray emission is generally limb-brightened,
but it is difficult to know the precise geometry of the
X-ray-emitting gas because extinction by dust within the
individual nebulae apparently severely distorts the X-ray maps.

As demonstrated by the models of \citet{zhekovperinotto1996}, this
gas should also emit in collisionally-excited, \lq\lq coronal"
lines of highly ionized species.  The two brightest coronal
emission lines in the optical range are \fex\ and \fexiv\
\citep{zhekovperinotto1996}.  If detected in planetary nebulae, a
measurement of their intensities would provide valuable
information about the physical properties of the emitting plasma.

In this paper, we present observations of \fex\ and \fexiv\ in
\bd, NGC 6543, NGC 7009, and NGC 7027.  All of these objects have been found to 
contain diffuse X-rays with either the Chandra X-ray Observatory or
XMM-Newton.  Throughout, we adopt wavelengths of 6374.51\AA\ and
5302.86\AA\ for \fex\ and \fexiv, respectively
\citep{trabert2004}. The observations and reductions are described
in Sect. 2. The results, null detections, are presented in Sect.
3. A discussion of the implications follows in Sect. 4 and Sect. 5.  Conclusions 
are presented in Sect. 6.

\section{Observations and Reductions}

\begin{figure*}
\includegraphics[width=16.5cm]{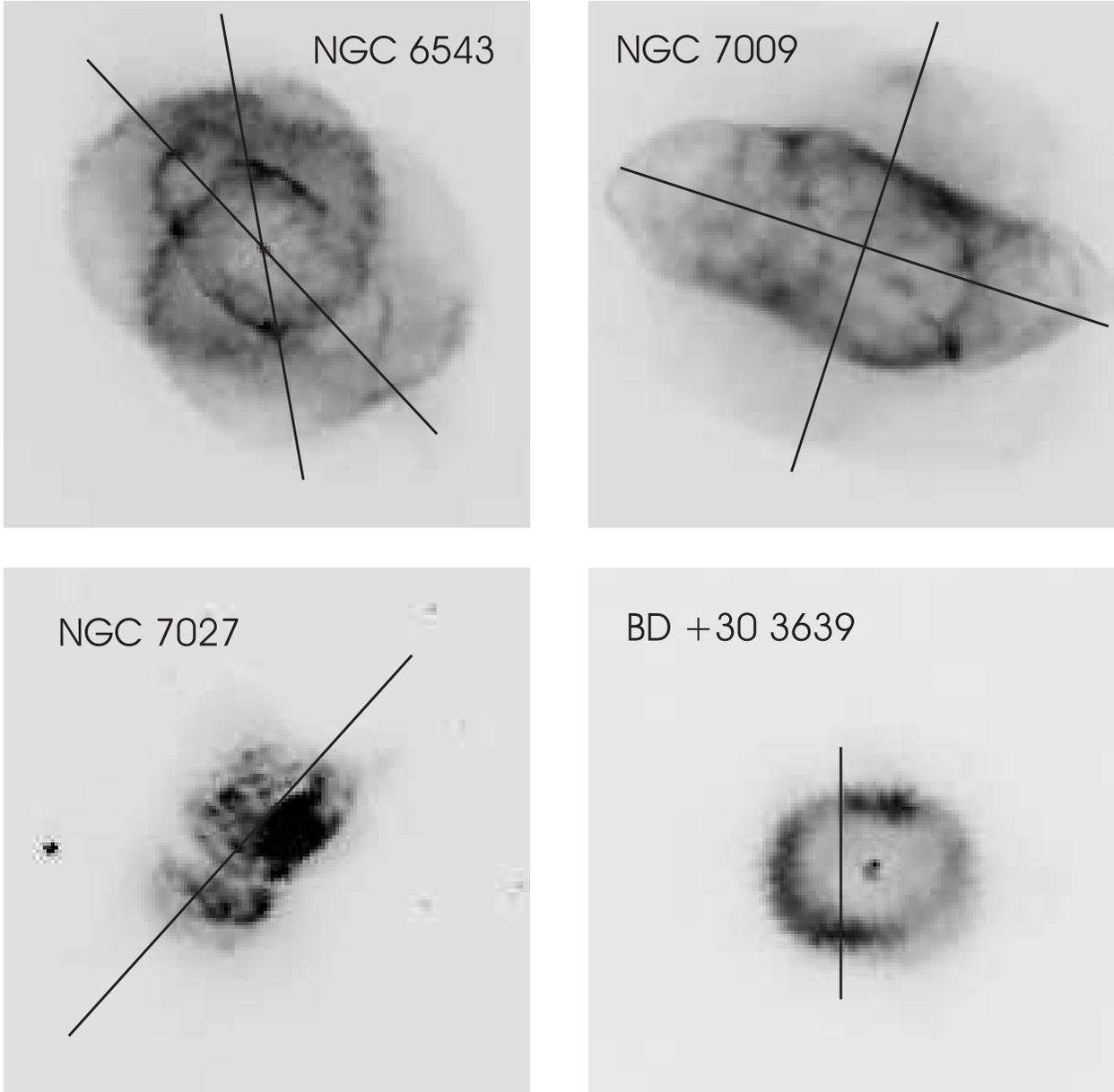}
\caption{HST images of the four objects studied in this paper (filter F656N for 
NGC 6543, NGC 7009, and \bd; filter F547M for NGC 7027). The
slit positions of our observations are superposed.  The slit length
is shown to scale, 6.\arcsec5 for \bd, 26\arcsec\ for the other objects.
North is up and East is left
in all of the images. } \label{all_obj}
\end{figure*}

Our observations were obtained on 2004 May 26-27 and 2004 August
2-4 (UT) with the REOSC echelle spectrograph on the 2.1-m
telescope of the Observatorio Astron\'omico Nacional in San Pedro
M\'artir, Baja California, Mexico (SPM). This is a conventional,
cross-dispersed echelle spectrograph whose dispersion is
0.190\,\AA/pix and 0.228\,\AA/pix at 5303\AA\ and 6374\AA,
respectively (a resolution, $\lambda/\delta\lambda$, of about $10^4$ 
for the 2.5-2.7 pix FWHM of arc lines).For observations of the standard stars, the 
slit width was 11\arcsec. The slit was oriented so as to cover various regions of
interest within each object (see Fig.
\ref{all_obj})\,\footnote{The images used in Fig.~\ref{all_obj}
are based upon observations made with the NASA/ESA Hubble Space
Telescope, obtained from the data archive at the Space Telescope
Science Institute. STScI is operated by the Association of
Universities for Research in Astronomy, Inc. under NASA contract
NAS~5-26555}.  In the case of NGC 7027, the position angle was chosen 
to approximately coincide with the position angle of the X-ray emission and was centered 
on the bright optical spot of the nebula. 
A GG395 filter was used to remove second order
contamination for \fex. The detector was a $1024 \times 1024$ SITe CCD with
24$\mu$m pixels. In the gain setting used, the CCD's readnoise was
approximately 9 electrons. The spectrograph was configured so that
emission from H$\alpha$ did not fall on the detector.
Nonetheless, some of its scattered light unavoidably contaminated
the adjacent order containing \fex.  Table \ref{obslog} presents a
log of the observations.

The data were reduced with IRAF\,\footnote{IRAF is distributed by the National Optical 
Astronomy Observatories, which are operated by the Association of Universities for Research 
in Astronomy, Inc., under cooperative agreement with the National Science Foundation.}.  
The initial step consisted of cleaning the spectra of the planetary nebulae of cosmic rays. 
Each night, bias images were obtained and used to construct a zero correction image that was 
subtracted from all exposures. Observations of standard stars were used to trace the positions 
and shapes of each order. For the planetary nebulae, these aperture definitions were shifted 
to the appropriate positions within the slit without changing the aperture shapes. When 
extracting the spectra of the standard stars, the inter-order background resulting from 
scattered light was subtracted.  This step was omitted for the spectra of the planetary 
nebulae since the scattered light was negligible.  The sky emission was not subtracted, 
but this poses no problem for our purposes. Spectra of a thorium-argon lamp were obtained 
as needed for wavelength calibration. The spectra of the standard stars were used to determine 
the instrumental sensitivity function within each order to provide a flux calibration. The 
atmospheric extinction was corrected using the published extinction curve for SPM 
\citep{schusterparrao2001}.

\section{Results}

\subsection{Upper limits to the intensities of \fex\ and \fexiv}

\begin{figure*}
\includegraphics[width=8cm]{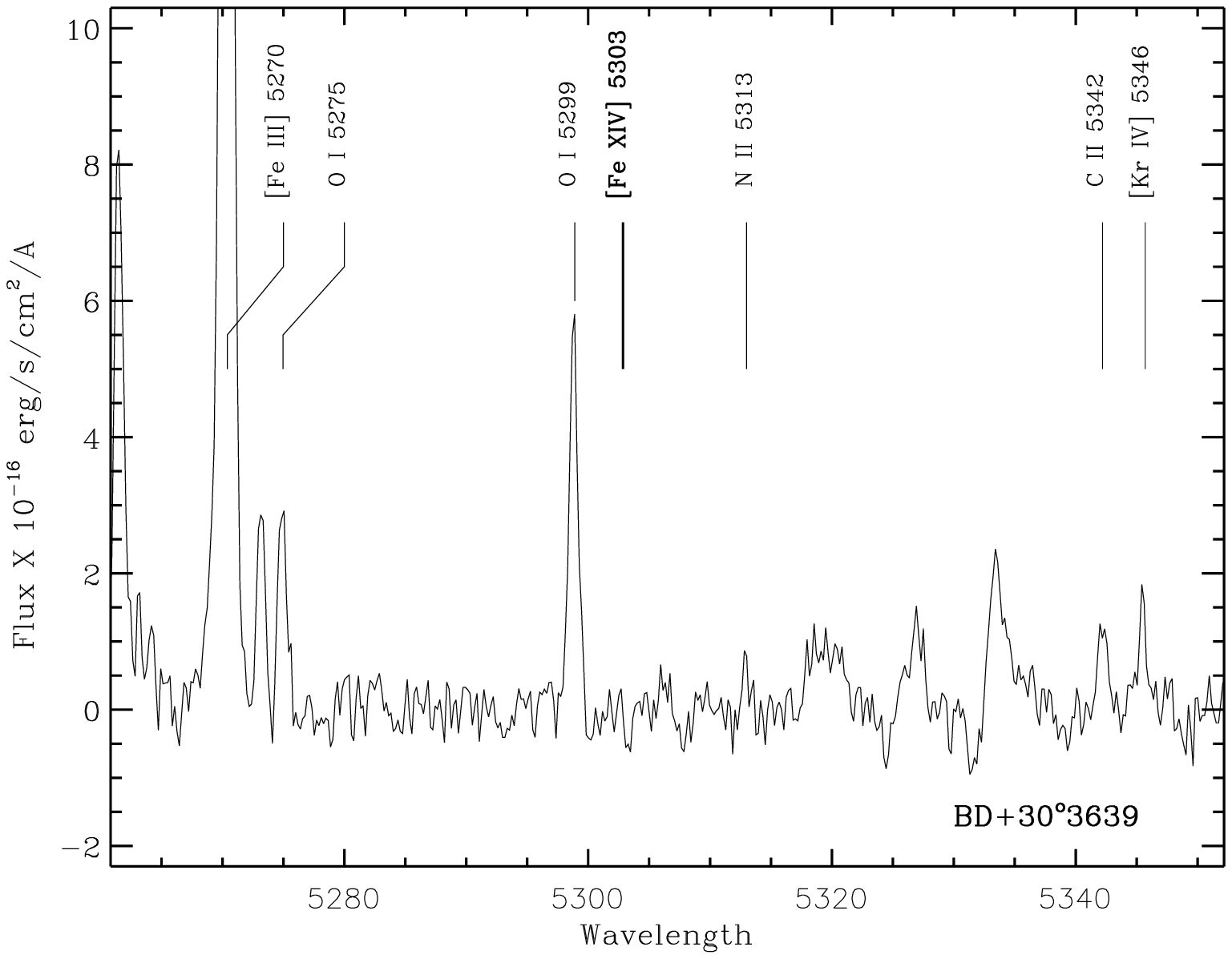} \ \ \
\includegraphics[width=8cm]{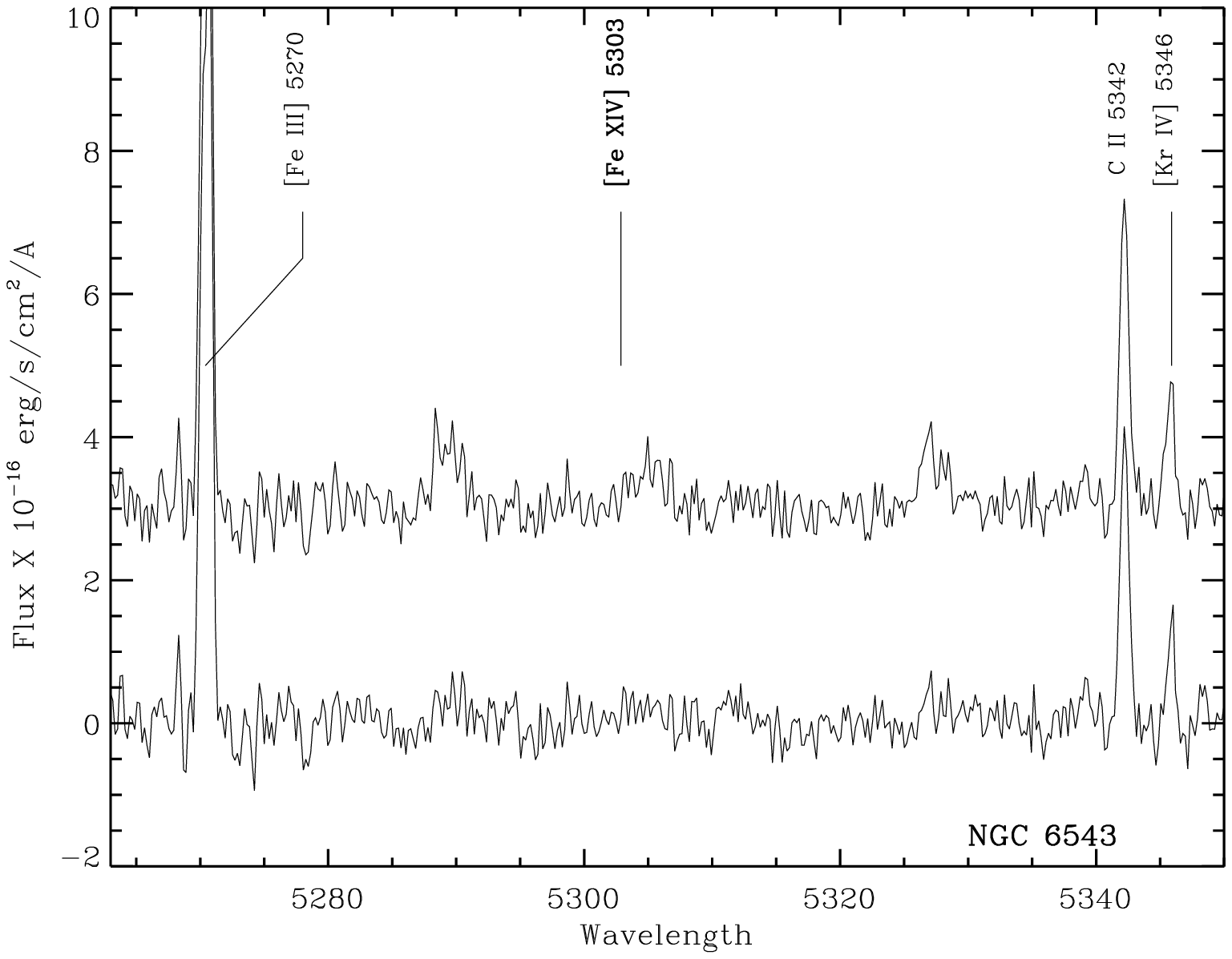}
\includegraphics[width=8cm]{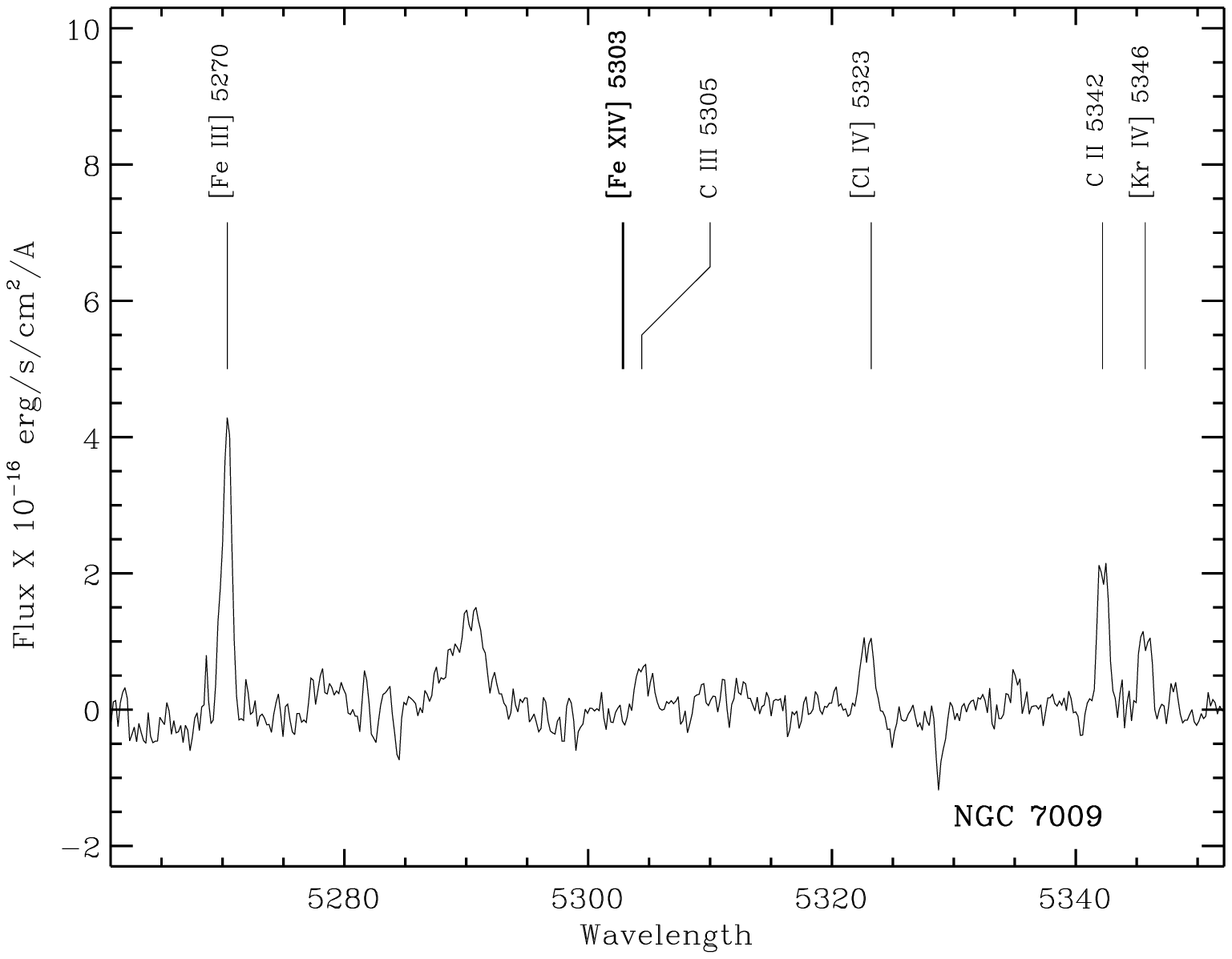} \ \ \
\includegraphics[width=8cm]{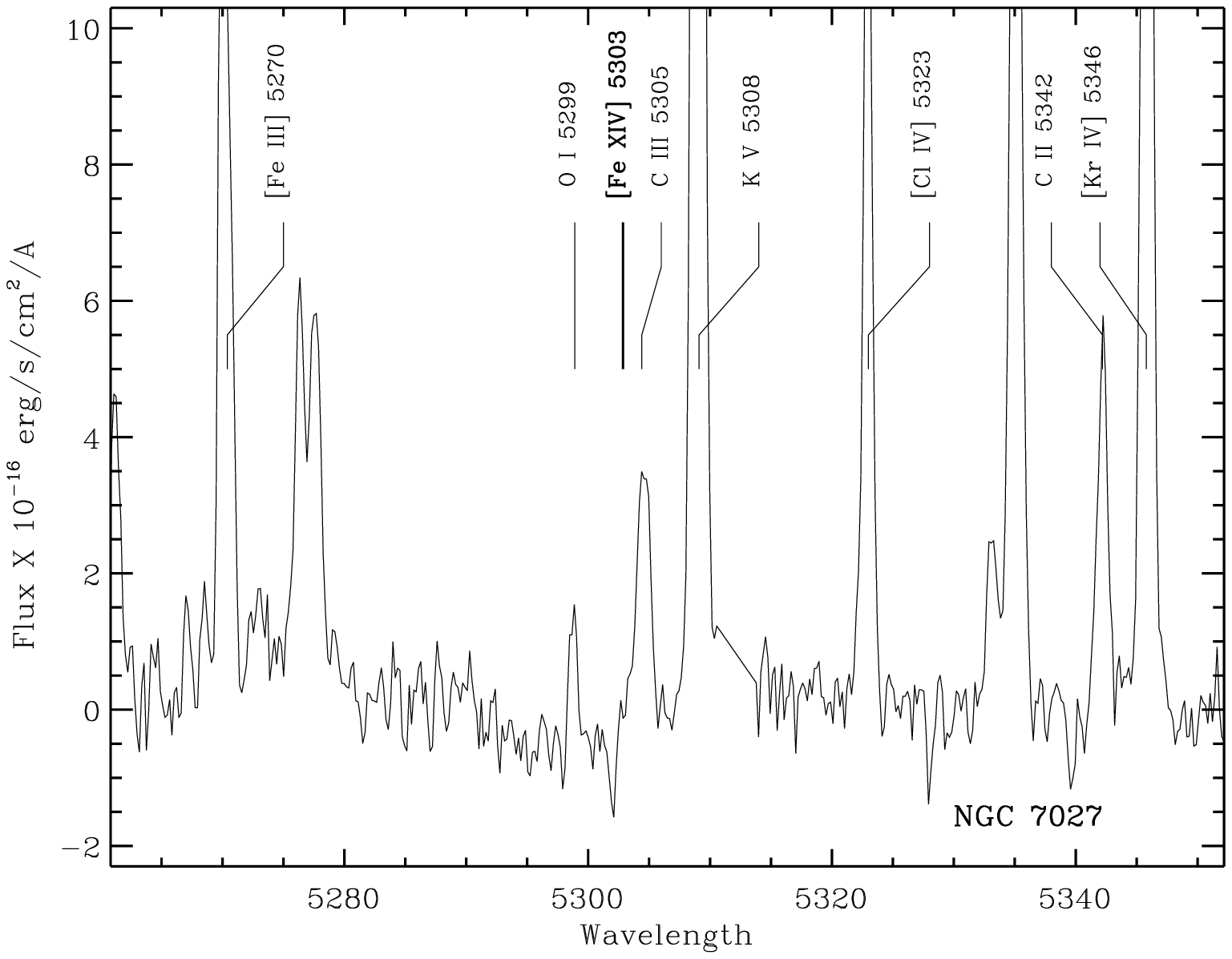}
\caption{We present our spectra of the region around \fexiv. The
spectra have been corrected to zero velocity and all continua have
been fit and subtracted. The position of \fexiv\ is indicated in
each figure.  Clearly, \fexiv\ is not present in any spectrum.
For NGC 6543, the top spectrum is the extraction including the
area of the slit occupied by the central star while the bottom
spectrum excludes this area, illustrating the removal of the
broad stellar emission.  } \label{figure_spec}
\end{figure*}

Particular care is required in the search for \fex\ and \fexiv.
The \fex\ line almost exactly coincides with emission from
He~I$\lambda$3187 in second order, so an order sorting filter is
absolutely essential.  Given the faintness of the \fex\ line and
the abundance of other very faint lines, such as O~{\sc
i}$\lambda$6374.32, only 0.19\AA\ distant from \fex, a spectral
resolution of 15,000-20,000 is required to assure sufficiently
precise wavelength calibration and line identification.  Such
resolution is normally achieved via echelle spectroscopy.  If
H$\alpha$ is found in the order adjacent to the \fex\ line, as is
the case with our observations, scattered light will be a
significant problem, even if H$\alpha$ is not included in the
echellogram.  Ideally, the light entering the spectrograph should
be filtered so as to exclude strong lines since they can be
sources of scattered light. For \fexiv, there are no problems with
either second order contamination or bright lines in adjacent
orders, but a high dispersion is nonetheless necessary to
discriminate it from other faint lines, e.g., C~{\sc
iii}$\lambda\lambda$5304.54, 5304.93, 5305.53 (see Fig. \ref{figure_spec}). Finally, when the
central star is visible, it is necessary to consider the emission
observed when the slit area including the central star is both
included and excluded since faint, broad stellar lines may also
cause confusion.

To help minimize confusion between \fex\ or \fexiv\ and other
faint lines, we corrected the spectra of each object to zero
velocity using other lines observed in the same order as the lines
of interest. In the case of \fexiv, we used the lines of
[Cl~IV]$\lambda$5323.3, [Fe~II]$\lambda$5333.7,
[Fe~VI]$\lambda$5335.3, C~II$\lambda$5342.5, and
[Kr~IV]$\lambda$5346.0, e.g., \citet{pequignotbaluteau1994}, all of
which appear in the same order.  In some cases, a faint line is
detected near the wavelength expected for \fexiv\ (Fig. \ref{figure_spec}), but its
wavelength is redshifted and coincides with that expected for
C~III$\lambda\lambda$5304.54, 5304.93, 5305.53.  It is clear that
the lines near \fexiv\ belong to C~III, since they are present in
the three objects in which He~II lines are also observed, but not
in \bd\ where He~II is absent. In the case of \fex, we used the
nearby line of Si~II$\lambda$6371.4 to define the zero point of
the velocity scale.

To further decrease confusion from other small signals, for all objects, we fit and
subtracted the stellar and nebular continua from each line of the 
two-dimensional spectra along the dispersion axis. 
In the case of \bd, the very strong, broad, stellar WR emission
feature resulting from the C~III lines coincides with \fexiv\
\citep[c.f.,][]{ackerneiner2003}.  This stellar emission feature was
fit as part of the continuum and subtracted.  The stellar C~III emission
seen in the top spectrum of NGC 6543 in Fig. \ref{figure_spec} is much,
much weaker than that in the spectrum of \bd\ and was not subtracted
because it would not obscure any significant \fexiv\ emission.
We then summed the corrected spectra along the spatial axis,
including and excluding the slit area occupied by the central
star, before attempting to measure the line intensities. Even so,
definite detections were not found and only upper limits could be
derived. We derived one sigma upper limits by computing the standard 
deviation of the noise about the mean continuum level.  We then obtained the
upper limits of the line intensities multiplying 
this noise amplitude by the Gaussian width of a nearby line.

The intensity limits observed for \fexiv, $F_{obs}(5303)$, may be found in Table
\ref{table_obsint}.  In this table, we present a single line intensity limit for 
NGC 6543 and NGC 7009 since the limits observed in both slit positions were similar.  
Fig.~\ref{figure_spec} presents continuum-subtracted portions of the spectra in the vicinity of
\fexiv\ in all of the objects observed.  The spectra in Fig. \ref{figure_spec} are integrated 
over the entire slit, except in the case of NGC 6543, where the sums including and excluding the central 
star are shown separately.  The position of \fexiv\ is indicated by a vertical bar and it is
clear that \fexiv\ was not detected in any object, though C~III$\lambda\lambda$5304.54,5304.93,5305.53 are, 
as previously noted.  We present neither spectra nor line intensity limits for \fex\ since this
line was never detected and the noise limits are similar to those
for \fexiv.  Since \fex\ is expected to be about 100 times fainter
than \fexiv\ (Section \ref{section_expint}), our flux limits for
\fex\ do not provide us with any additional information.

Table \ref{table_obsint} also presents reddening-corrected line intensity limits for later comparison with model predictions of the \fexiv\ intensity. The reddening-corrected flux limits, $F_{RC}(5303)$,
are calculated from the observed flux limits, $F_{obs}(5303)$,
according to
\begin{equation}
F_{RC}(5303) = F_{obs}(5303)\times 10^{0.4 E(B-V)
A(5303)} \label{eq_reddening}
\end{equation}
where $E(B-V)$ is the reddening and $A(5303) = 3.10$\,mag is the
\citet{fitzpatrick1999} reddening value, parametrized with a value
of the total-to-selective extinction of 3.041
\citep{mccallarmour2000}.  For \bd\ and NGC 7027, where there are significant, small scale variations in reddening \citep{kastneretal2002}, we provide the range of reddenings within our slit and the corresponding reddening-corrected fluxes as ranges in Table \ref{table_obsint}.

\begin{deluxetable}{lccc}
\tablecolumns{4} \tablecaption{Upper limits to the observed intensities of \fexiv}
\tablehead{ \colhead{Object} & \colhead{$F_{obs}(5303)$} & \colhead{$E(B-V)$} & \colhead{$F_{RC}(5303)$} \\ 
\colhead{} & \colhead{erg\,s$^{-1}$\,cm$^{-2}$} & \colhead{mag} & \colhead{erg\,s$^{-1}$\,cm$^{-2}$} }
\startdata
\bd      & $< 5.4\times 10^{-17}$ & 0.49 [0.23-0.66] & $< 2.2\times 10^{-16}$ [$1.0-3.5\times 10^{-16}$]\\
NGC 6543 & $< 5.3\times 10^{-17}$ & 0.21 & $< 9.5\times 10^{-17}$ \\
NGC 7009 & $< 1.8\times 10^{-17}$ & 0.11 & $< 4.6\times 10^{-17}$ \\
NGC 7027 & $< 5.6\times 10^{-17}$ & 0.99 [0.49-1.97]& $< 9.3\times 10^{-16}$ [$2.3-155\times 10^{-16}$]\\
\enddata
\label{table_obsint}
\end{deluxetable}

\subsection{Expected intensities of \fex\ and
\fexiv}\label{section_expint}

Since we detected neither \fex\ nor \fexiv\ in any of the objects we
observed, it is interesting to compute the intensities expected for these lines.  Our estimate is based upon the observed X-ray spectrum since the hot bubble that emits the X-rays is the source for coronal line emission as well.  We adopted the following procedure to compute the intensity of \fex\ and \fexiv.  First, from the X-ray spectrum {\bf extracted for the entire object}, we derived the plasma parameters  (temperature, abundances, and the emission measure) required to explain its X-ray emission.  Second, we re-scaled the emission measure to the size of the slit used for our optical observations.  Finally, based upon the emission measure \lq\lq in the slit" and the plasma
temperature of the hot gas, we estimated the emission for \fex\ and \fexiv\ using the same optically-thin plasma model that was used to derive the plasma parameters in the first step.

All the necessary X-ray data were obtained from the {\it Chandra} (NGC 6543: \dataset[ADS/Sa.CXO\#obs/00630]{Chandra ObsId 630}; BD$+$303639: \dataset[ADS/Sa.CXO\#obs/00587]{Chandra ObsId 587}; and NGC 7027: \dataset[ADS/Sa.CXO\#obs/00588]{Chandra ObsId 588}) and {\it XMM-Newton} (NGC 7009) science archives. The Chandra observations were carried out with the ACIS-S detector and the corresponding Science Threads for Image Spectroscopy in CIAO 3.2\,\footnote{Chandra Interactive Analysis of Observations (CIAO),
http://cxc.harvard.edu/ciao/} were used for extracting the X-ray spectra.  For the XMM-Newton observation of NGC 7009, the most recent version of SAS 6.1.0\,\footnote{XMM-Newton Science Analysis System (SAS),
http://xmm.vilspa.esa.es/external/xmm$\_$sw$\_$cal/sas$\_$frame.shtml} was used for the spectral extraction.  Given the goal of our study, only the EPIC PN spectrum was used in the following analysis due to its better photon statistics compared to those of the EPIC MOS1 and MOS2. The periods of high background count rates were excluded from the EPIC PN data. Finally, for each object, X-ray spectra were extracted corresponding to the entire object and that part within the optical slit.  The X-ray spectra were fit within XSPEC 11.3 \citep{arnaud1996} using the {\it vapec} model for the emission from optically-thin plasma. 

\begin{figure*}
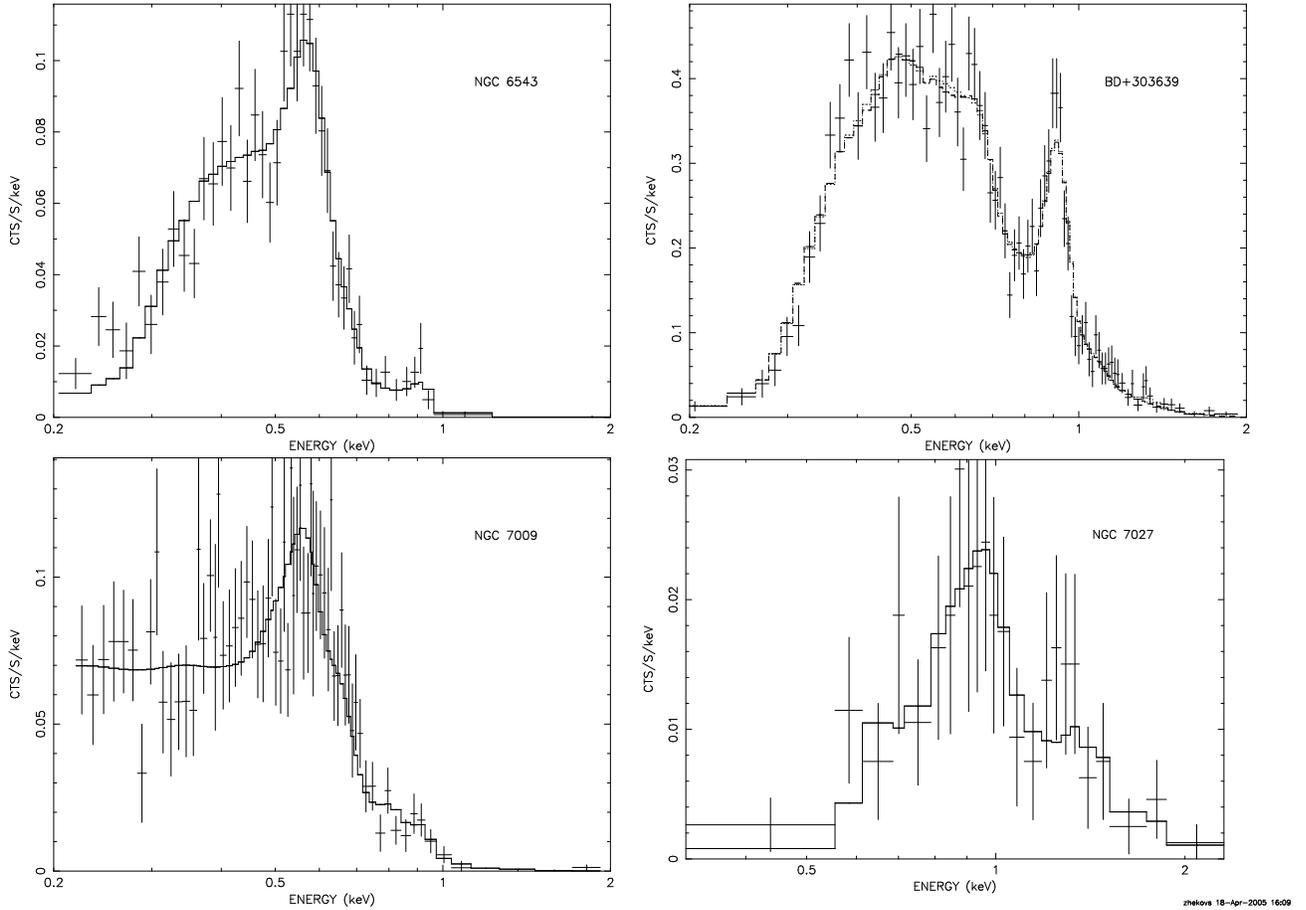

\includegraphics[width=6cm,angle=-90]{f3a} \ \ \
\includegraphics[width=6cm,angle=-90]{f3b}
\includegraphics[width=6cm,angle=-90]{f3c} \ \ \
\includegraphics[width=6cm,angle=-90]{f3d}
\caption{Our fits to the X-ray spectra are based upon the parameters in Tables \ref{table_xrayfit3} and \ref{table_7027par}.  The two fits for \bd\ use the two abundances sets given in Table \ref{table_xrayfit3}. Note that the two model
spectra (dashed and dotted lines) overlap almost perfectly 
(except around $\sim$ 0.6 keV).
For NGC 7027, the fit shown is that for the parameters in the last column of Table \ref{table_7027par}.  } \label{figure_xrayspec}
\end{figure*}

We note that the derived parameters of the X-ray 
emitting plasma do not depend on this choice and using a different model 
(e.g. {\it vmekal}) leads to the same results within the expected 
uncertainties\footnote{For each data set, we performed various tests 
     by fitting the spectra both with {\it vapec} and {\it vmekal}.
     The differences between the derived plasma temperatures and X-ray 
     absorptions from the both plasma models were less than 4\%, and
     the abundances overlap within the corresponding 90\% confidence
     limits. General comments on such comaprisons can be found in
     http://cxc.harvard.edu/atomdb/issues$\_$comparisons.html}
.  Thus, the {\it vapec} model was chosen to make calculations of the
\fex\ and \fexiv\ emissions technically more straightforward.
The hot plasma abundances were taken from \citet{manessvrtilek2003} for NGC 6543, from \citet{manessetal2003} for NGC 7027, and from \citet{guerreroetal2002} for NGC 7009. For \bd, two abundance sets were used, the first being a variant of the \citet{manessetal2003} abundance set (fourth column in Table \ref{table_xrayfit3}; henceforth, we denote this set as `wind' abundances) 
and the second a new set derived here (last column in Table \ref{table_xrayfit3}; `nebular' abundances) based partly upon the nebular abundances 
of \citet{allerhyung1995} and which have values similar to the 
`nebular' abundances derived by \citet{arnaud1996a} from the analysis of the
ASCA data of this object.  The two fits are indistinguishable in fitting the data.  Figure \ref{figure_xrayspec} presents our fits to the X-ray spectra.  The parameters used in these fits, including elemental abundances, are listed in Tables \ref{table_xrayfit3} and \ref{table_7027par}.  Our fits to the total X-ray spectra of these PNe are consistent with those obtained by other groups using the same data sets.  Finally, we note that, if the \citet{gorenstein1975} conversion is used, $N_H = 2.22\times 10^{21} A_V$~cm$^{-2}$, the column densities derived for the X-ray absorption are consistent with those from the optical extinction.

\begin{figure*}
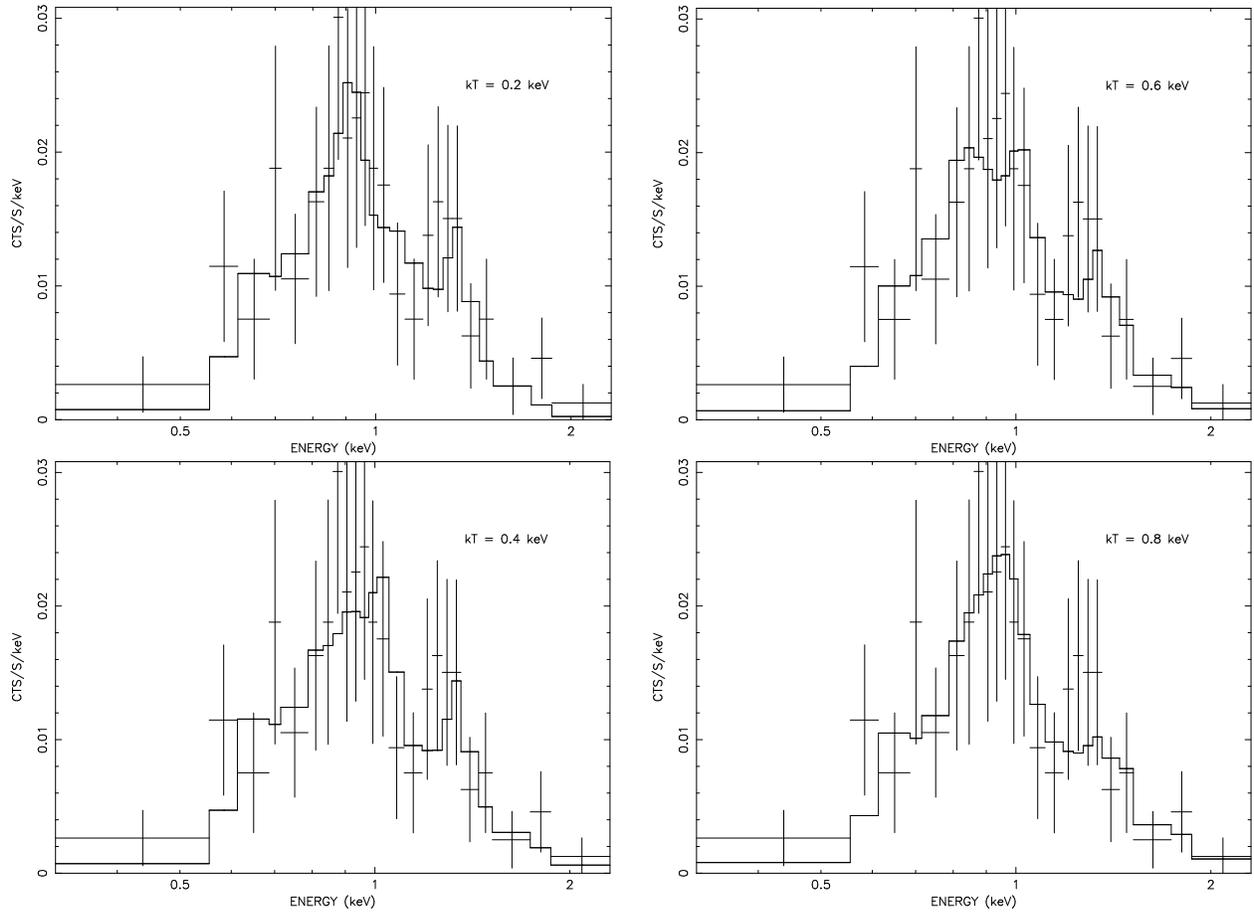

\includegraphics[width=6cm,angle=-90]{f4a} \ \ \
\includegraphics[width=6cm,angle=-90]{f4b}
\includegraphics[width=6cm,angle=-90]{f4c} \ \ \
\includegraphics[width=6cm,angle=-90]{f4d}
\caption{These fits to the X-ray spectrum of NGC 7027 demonstrate how poorly this spectrum constrains the fit parameters (see Table \ref{table_7027par}).  The parameters used here bracket those available in the literature \citep{kastneretal2001,manessetal2003}.} \label{figure_7027par}
\end{figure*}

Before proceeding to the predicted \fex\ and \fexiv\ fluxes, a few comments are necessary regarding \bd\ and NGC 7027.  As noted above, the two abundance sets used for \bd\ produce fits to the total X-ray spectrum that are equally good (Table \ref{table_xrayfit3}, Fig. \ref{figure_xrayspec}).  This clearly demonstrates the limitations of attempting to constrain the hot-plasma abundances in this object from the available data.  
    In contrast to \citet{manessetal2003}, our `wind' abundance set uses
    an Fe abundance of 1.0 with respect to solar, rather than an $\mathrm {Fe}$
    abundance of 0.0, chosen simply to illustrate that the current
    data do not constrain the iron abundance in this case. It fact, \citet{manessetal2003} 
    did not derive the iron abundance, but postulated
    it equal to 0.0 with respect to solar). On the other hand, an
    iron abundance of 0.0 is derived only if `nebular' abundances are
    used, and, in this case, the solar value of 1.0 for iron abundance 
    is incompatible with the data, as the last column in \ref{table_xrayfit3} makes 
    clear.
So, this experiment demonstrates that the spectrum may be fit using plasma with iron abundances between zero and the solar value.  As for NGC 7027, our experiments in fitting its X-ray spectrum demonstrate the limitations of the low counting statistics in this spectrum (Table \ref{table_7027par}, Fig. \ref{figure_7027par}).  The plasma temperature was set at four steps to explore the range of values from previous analyses
(kT = 0.26 keV \citet{kastneretal2001}; kT = 0.72 keV \citet{manessetal2003}) and the spectrum fit with a lower number of free parameters.  We finally adopted the fit presented in the last column of Table \ref{table_7027par}, since these parameters nominally produce the best fit.  

\begin{deluxetable}{lllll}
\tablecolumns{5} 
\tablecaption{X-ray fit results for NGC 6543, NGC 7009, and \bd\ $^a$}
\tablehead{ \colhead{} & \colhead{NGC 6543} & \colhead{NGC 7009} & 
\colhead{\bd$^{f}$} & \colhead{\bd$^{g}$} }
\startdata
Counts\ $^{\mathrm b}$ & 1551 & 1891 & 4569 & 4569 \\
Exposure (ks) & 46.0 & 31.5\ $^{\mathrm c}$ & 18.9 & 18.9 \\
\hline
$\chi^2$/dof  & 42.4/39 &  62.8/65   &  86.8/73 & 83.7/71 \\
N$_H$(10$^{21}$ cm$^{-2}$) & 0.0 [0.0 - 0.4] & 0.38 [0.19 - 0.62] & 2.0 [1.7 - 2.3] & 2.0 [1.5 - 2.2] \\
kT (keV)  & 0.15 [0.14 - 0.16] & 0.19 [0.18 - 0.20] & 0.20 [0.18 - 0.22] & 0.19 [0.15 - 0.22] \\
H    & 1 & 1  & 1 & 1 \\
He   & 1 &  1.33   & 1 & 1 \\
C    & 0.69 &  0.8   & 348 [101 - 1000] & 3.70 [0.39 - 1000] \\
N    & 1.40 [0.68 - 2.06] &  1.9  & 20.8 [15.8 - 98.5] & [0.15 [0.02 - 71.7] \\
O    & 0.38 [0.26 - 0.52] & 0.44 [0.33 - 0.58]  & 3.6 [1.23 - 12.4] & 0.05 [0.02 - 11.6] \\
Ne   & 1.08 [0.44 - 2.15] &  0.77 [ 0.35 - 1.28]  & 19.3 [6.6 - 71.0] & 0.32 [0.14 -69.1] \\
Mg   & 1 &   1  & 1 & 0.15 [0.0 - 69.1] \\
Si   & 1  & 1  & 1 & 1 \\
S    & 0.56  & 1.2 & 1 & 0.33 \\
Ar   & 1.55  &  1  & 1 & 0.05 \\
Ca   & 1  &  1  & 1 & 1 \\
Fe   & 1  &  1  & 1 & 0.0 [0.0 - 0.29] \\
Ni   & 1  &  1  & 1 & 1 \\
norm$^{d}$   &  $1.49\times 10^{10}$ & $1.11\times10^{10}$ & $1.39\times 10^{10}$ & $9.48\times 10^{11}$ \\
F$_X$(0.3-2 keV)$^e$ & 1.05 &  0.82 & 6.41 & 6.49 \\
\enddata
\tablenotetext{a}{The 90\% confidence limits for the fit parameters are given in brackets.  All abundances are with respect to the solar values of \citet{andersgrevesse1989}.}
\tablenotetext{b}{This is the number of counts in the {\it total} X-ray spectrum.}
\tablenotetext{c}{This is the effective exposure time for the EPIC PN detector after applying a correction for the high background events.}
\tablenotetext{d}{norm $= EM/(4 \pi d^2)$~cm$^{-5}$, where $EM = \int n_e n_H dV$ is the emission measure of the hot gas and $d$ is the distance to the object.}
\tablenotetext{e}{The {\it observed} X-ray flux is in units of $10^{-13}$ ergs cm$^{-2}$ s$^{-1}$.}
\tablenotetext{f}{`Wind' abundances.}
\tablenotetext{g}{`Nebular' abundances.}
\label{table_xrayfit3}
\end{deluxetable}

\begin{deluxetable}{lllll}
\tablecolumns{5} 
\tablecaption{X-ray fit results for NGC 7027 $^a$}
\tablehead{ \colhead{} & \colhead{NGC 7027} & \colhead{NGC 7027} & \colhead{NGC 7027} & \colhead{NGC 7027} }
\startdata
Counts\ $^{\mathrm b}$ & 278 & 278 & 278 & 278 \\
Exposure (ks) & 18.2 & 18.2 & 18.2 & 18.2 \\
\hline
$\chi^2$/dof  & 11.1/19 & 11.1/19  & 10.3/19  & 9.0/19    \\
N$_H$(10$^{21}$ cm$^{-2}$) & 10.4 [8.3 - 12.4] & 6.4 [3.7 - 8.5]  &
                             5.0 [0. - 7.4]  & 4.3 [0. - 6.1] \\
kT (keV)  & 0.2  & 0.4  & 0.6 & 0.8  \\
H    & 1 & 1 & 1  & 1 \\
He   & 1.08 & 1.08 & 1.08  & 1.08 \\
C    & 1.69 & 1.69  & 1.69 & 1.69 \\
N    & 1.71  & 1.71  & 1.71 & 1.71 \\
O    & 3.9 [0.45 - 96.6] & 10.0 [1.4 - 48.0] &  
                       8.8 [1.1 - 36.8] & 10.0 [0.8 - 28.9] \\
Ne   & 2.7 [0.44 - 1000] & 5.9 [1.2 - 896] &  
                       3.9 [0. - 13.5] & 0. [0. - 9.7] \\
Mg   & 3.6 [0 - 1000] & 4.8 [0. - 869] &  
                       3.4 [0. - 29.8] & 2.2 [0. - 11.7] \\
Si   & 0.17  & 0.17 &  0.17 & 0.17 \\
S    & 0.51  & 0.51 & 0.51 & 0.51 \\
Ar   & 0.63  & 0.63 & 0.63  & 0.63 \\
Ca   & 1  &  1 &  1 & 1 \\
Fe   & 1  &  1  & 1 & 1 \\
Ni   & 1  &  1 & 1  & 1 \\
norm$^{c}$   &  $9.86\times10^{10}$ & $4.64\times10^{9}$  &
                      $3.65\times10^{9}$ & $3.72\times10^{9}$ \\
F$_X$(0.3-2 keV)$^d$ & 0.37 & 0.38  &  0.38 & 0.39  \\
\enddata
\tablenotetext{a}{The 90\% confidence limits for the fit parameters are given in brackets.  All abundances are with respect to the solar values of \citet{andersgrevesse1989}.}
\tablenotetext{b}{This is the number of counts in the {\it total} X-ray spectrum.}
\tablenotetext{c}{norm $= EM/(4 \pi d^2)$~cm$^{-5}$, where $EM = \int n_e n_H dV$ is the emission measure of the hot gas and $d$ is the distance to the object.}
\tablenotetext{d}{The {\it observed} X-ray flux is in units of $10^{-13}$ ergs cm$^{-2}$ s$^{-1}$.}
\label{table_7027par}
\end{deluxetable}

The fluxes in \fex\ and \fexiv\ predicted for the four objects under study are given in Table \ref{table_predint}.  The model \fexiv\ flux is linearly proportional to the iron abundance and all of the predicted fluxes are based upon models with a solar iron abundance.  The same plasma model that was used to fit the X-ray spectra was used for estimating the fluxes of coronal line emission for \fex\ and \fexiv, the only difference being to re-scale the emission measure to account for the fraction of the object included in the optical spectroscopy.  The scaling procedure was internally checked for consistency by (i) comparing the number of photons in the total and \lq\lq slit" X-ray spectra and (ii) fixing the model parameters to their values from the fit to the total spectrum and varying only the normalization.  Both approaches gave identical results (identical spectral normalization parameter). An important advantage of estimating the coronal line emission using the fit to the X-ray spectrum is that uncertainties in the distance to object disappear, due to the definition of the normalization parameter for the fit to the X-ray spectrum, $norm = EM/(4 \pi d^2)$~cm$^{-5}$, $EM = \int n_e n_H dV$~ being the emission measure of the hot gas and $d$ the distance to the object.  In this way, the predicted X-ray spectrum and the predicted flux in coronal lines are internally consistent.  The fluxes in Table \ref{table_predint} correspond to what should be observed using the slits employed for the optical observations and may be compared directly to the reddening-corrected limits observed in Table \ref{table_obsint}.  

It is immediately evident that there are severe discrepancies in the \fexiv\ flux limits observed and the fluxes predicted by models for NGC 7009, NGC 6543, and \bd, with the observed limits lower by factors of 3.5, 12, and 8, respectively.  For NGC 7027, there is only a discrepancy if the plasma temperature has the lowest value of the four considered in Table \ref{table_7027par}.    

The predicted fluxes for \fex\ explain immediately why this line was never detected.  The predicted intensity of this line is often fainter than \fexiv\ by 2 orders of magnitude or more.  As a result, it is always expected to be below our detection limits.  

\begin{deluxetable}{ccccc}
\tablecolumns{5} 
\tablecaption{Predicted fluxes of \fex\ and \fexiv\ $^{\mathrm a}$}
\tablehead{ \colhead{Object} & \colhead{PA} & \colhead{kT} & \colhead{\fex} & \colhead{\fexiv} \\
\colhead{} & \colhead{deg} & \colhead{keV} & \colhead{erg\,s$^{-1}$\,cm$^{-2}$} & \colhead{erg\,s$^{-1}$\,cm$^{-2}$} }
\startdata
NGC 6543       &  10  &  0.15 & $2.17\times10^{-17}$ & $1.42\times10^{-15}$ \\
NGC 6543       &  43  &  0.15 & $1.72\times10^{-17}$ & $1.13\times10^{-15}$ \\
\bd            &   0  &  0.20 & $5.55\times10^{-19}$ & $1.77\times10^{-15}$ \\
NGC 7009       &  72  &  0.19 & $1.05\times10^{-19}$ & $1.60\times10^{-16}$ \\
NGC 7009       & 162  &  0.19 & $1.26\times10^{-19}$ & $1.93\times10^{-16}$ \\
NGC 7027       & 138  &  0.2 & $3.23\times10^{-18}$ & $1.03\times10^{-14}$ \\
NGC 7027       & 138  &  0.4 & $9.65\times10^{-23}$ & $2.58\times10^{-18}$ \\
NGC 7027       & 138  &  0.6 & 0. \hspace{1cm} & $6.49\times10^{-20}$ \\
NGC 7027       & 138  &  0.8 & 0. \hspace{1cm} & $2.21\times10^{-21}$ \\
\enddata
\tablenotetext{a}{The \fex\ and \fexiv\ fluxes assume a {\it solar} iron abundance (Fe $= 1$) in Tables \ref{table_xrayfit3} and \ref{table_7027par}.}
\label{table_predint}
\end{deluxetable}

\section{The missing iron}

Comparing the observed limits from Table \ref{table_obsint} with the X-ray based predictions for \fexiv\ in Table \ref{table_predint}, it is abundantly clear that there is a large discrepancy.  The models consistently predict stronger \fexiv\ emission than is observed, by factors of at least a few to 12.  Only in NGC 7027 do the models based upon the X-ray spectrum predict \fexiv\ 
emission fainter than our limits, provided a higher plasma temperature is adopted 
in this object, but we recall that the plasma parameters are not well 
constrained due to the quality of its X-ray spectrum.
In the other three objects, all analyses to date have found much lower plasma temperatures, so this mechanism may be ruled out in those cases.  Considering all of the uncertainties from our analysis, we suggest that the most plausible explanation for this observed discrepancy is that the gas in the hot bubble in these three planetary nebulae is strongly {\it iron-depleted}. 

\bd\ provides an illustrative example.  Our experiments with abundance sets demonstrates that the X-ray spectrum does not meaningfully constrain the abundances in the X-ray-emitting gas (Table \ref{table_xrayfit3}).  The spectrum may be fit by abundance sets with and without iron, the other elements being adjusted in a plausible fashion, and approximately matching the abundances found for the nebular gas and the stellar atmosphere.  If an abundance set depleted in iron is adopted, the \fexiv\ flux predicted may be brought into agreement with observations, but this also serves to illustrate that it is inescapable that iron is depleted in the plasma emitting the X-rays in \bd.  Clearly, it would be extremely useful to determine both the atmospheric iron abundance of the central star as well as the nebular iron abundance, since these would set bounds on the abundances expected in the hot gas.  

\section{Implications for the origin of the X-ray gas}

The depletion of iron in the gas emitting in X-rays is a strong chemical signature that is of potential use in identifying the origin of this gas.  The gas emitting in X-rays should have the composition of either the nebular gas, the stellar wind, or some mixture of the two.  

Nebular iron abundances are rare.  However, they do exist for NGC 6543 and NGC 7027.  \citet{perinottoetal1999} found depletions by factors of 11 and 80 in NGC 6543 and NGC 7027, respectively, the extreme values they found for their sample of four planetary nebulae.  Similarly, \citet{sterlingetal2005} found a depletion of iron by a factor of 3-14 in SwSt-1 depending upon the region of the nebula observed.  It is therefore clear that iron may often be depleted by a factor of 10 or more in planetary nebulae, though more extensive studies would obviously be very helpful. The remaining fraction of the iron is presumably found in dust grains.  Similar depletion of iron onto dust grains is also found in H~{\sc ii} regions, e.g., \citet{rodriguez2002} found iron depletions by factors of 3 to 50. 

Unfortunately, nothing is known of the iron abundance in the stellar wind for the objects studied here.  \bd\ and NGC 6543 have hydrogen-deficient central stars, NGC 7009 has a hydrogen-rich central star, and nothing is known of the central star in NGC 7027.  Generally, there appears to be a clear trend that hydrogen-deficient central stars as well as their supposed PG-1159 and WO progeny, with or without winds, have atmospheres deficient in iron by typically at least an order of magnitude \citep{miksaetal2002, heraldbianchi2004a, heraldbianchi2004b, heraldbianchi2004c,stasinskaetal2004}.  This iron deficiency may be the result of nuclear processing \citep[e.g.,][]{herwigetal2003}.  On the other hand, the central stars of planetary nebulae with hydrogen-rich atmospheres that still have winds do not appear to be depleted in iron \citep{heraldbianchi2004a, heraldbianchi2004b, heraldbianchi2004c}.

The depletion of iron in either the stellar wind or nebular material could explain our failure to detect \fexiv.  Clearly, the depletion of iron onto dust grains in the nebular gas allows iron depletion in all objects, while the depletion due to atmospheric composition allows depletion only in \bd\ and NGC 6543.  That we find an iron deficiency irrespective of the properties of the central stars could imply that the depletion process is also independent of the central stars.  In this case, the hot gas should be of nebular origin.  That the X-ray spectra in all four objects may all be modelled using a plasma with the nebular abundances further reinforces the impression that the plasma emitting in X-rays has a nebular origin.

If the X-ray emitting plasma arises from nebular gas and the iron depletion is due to iron being locked up in dust grains, the dust must survive in the X-ray plasma.  Dust survival therefore probably excludes severe shocking of this plasma material \citep[e.g.,][]{tielensetal1994}.  It may be more likely that this matter is heated by a more gentle mechanism, such as heat conduction.  The work of \citet{tielensetal1994} indicates that iron grains should survive sputtering from thermal ions over the lifetime of a planetary nebula, provided that the original grains are reasonably large, $0.1-1.0\,\mu$m.

Finally, although our finding of iron depletion in the gas emitting in X-rays offers a chemical signature with the potential to discern the origin of this gas, at present it is impossible to draw a firm conclusion.  To progress will require measuring the iron abundances in the nebular gas and the atmospheres of the central stars of these planetary nebulae.  

\section{Conclusions}

We have attempted to observe the \fex\ and \fexiv\ emission from four planetary nebulae in which diffuse X-ray emission is observed.  The plasma emitting in X-rays should also emit in the coronal lines we observed, but our observations yield only upper limits for both lines in all four objects.  Based upon the X-ray spectra of the gas emitting in X-rays, we estimate the intensities expected for the \fex\ and \fexiv\ lines using the same models that reproduce the observed X-ray spectra.  \fex\ is expected to be much weaker than our observed upper limits, so it is understandable that we failed to detect it. For \fexiv, however, the expected intensities substantially exceed our observed one sigma upper limits, by factors of 3.5, 12, and 8 in NGC 7009, NGC 6543, and \bd, respectively.  In NGC 7027, the higher plasma temperature explains the absence of \fexiv\ emission.

The simplest explanation we find for this discrepancy is a depletion of iron in the plasma emitting the X-rays in all objects.  This depletion may be checked with further observations of the iron abundance in both the nebular gas and stellar atmospheres.  In principle, this chemical signature could be used to determine whether the gas emitting in X-rays arises from nebular gas or the stellar wind.  While a variety of evidence appears to favour a nebular origin for the X-ray-emitting gas, no definitive conclusion is possible given the chemical abundances currently available.  

\acknowledgments

We acknowledge the able assistance of Gabriel Garc\'\i a, Gustavo
Melgoza, and Felipe Montalvo with the observations. We acknowledge
financial support from CONACyT projects 34522-E, 37214-E, 40864-E, and 42809-E  and
DGAPA projects IN107202 and IN112103.  SAZ acknowledges financial support by NASA through  Chandra
Award G04-5072A.  We thank Jos\'e Alberto L\'opez, Gra\.zyna
Stasi\'nska and Gloria Koenigsberger for very useful discussions.

\end{document}